\documentclass[journal,twoside,web]{ieeecolor}
\usepackage{tmi}
\usepackage{cite}
\usepackage{amsmath,amssymb,amsfonts}
\usepackage{algorithmic}
\usepackage{graphicx}
\usepackage{multirow}
\usepackage{textcomp}
\usepackage{float}
\usepackage{epstopdf}
\usepackage{flushend}

\def\BibTeX{{\rm B\kern-.05em{\sc i\kern-.025em b}\kern-.08em
    T\kern-.1667em\lower.7ex\hbox{E}\kern-.125emX}}
\begin{document}
\title{Estimating dual-energy CT imaging from single-energy CT data with material decomposition convolutional neural network}
\author{Tianling Lyu, Zhan Wu, Yikun Zhang, Yang Chen, \IEEEmembership{Senior Member, IEEE}, Lei Xing, Wei Zhao
\thanks{This work was supported in part by the State ’s Key Project of Research and Development Plan under Grant 2017YFA0104302, Grant 2017YFC0109202 and 2017YFC0107900, in part by the National Natural Science Foundation under Grant 81530060 and 61871117, in part by Science and Technology Program of Guangdong (2018B030333001), in part by NIH (1R01CA223667 and R01CA227713) and a Faculty Research Award from Google Inc. (\textbf{Corresponding author: Yang Chen, Lei Xing and Wei Zhao})}
\thanks{T. Lyu is with the Laboratory of Image Science and Technology, Southeast University, Nanjing 210096, China, and also with the the Department of Radiation Oncology, Stanford University, Palo Alto, CA 94306, USA.}
\thanks{Z. Wu and Y. Zhang are with the Laboratory of Image Science and Technology, Southeast University, Nanjing 210096, China, and also with the Key Laboratory of Computer Network and Information Integration (Southeast University), Ministry of Education, Nanjing 210096, China.}
\thanks{Y. Chen is with the Laboratory of Image Science and Technology, Southeast University, Nanjing 210096, China, and also with the Key Laboratory of Computer Network and Information Integration (Southeast University), Ministry of Education, Nanjing 210096, China (e-mail: chenyang.list@seu. edu.cn).}
\thanks{L. Xing is with the Department of Radiation Oncology, Stanford University, Palo Alto, CA 94306, USA (e-mail: lei@stanford.edu).}
\thanks{W. Zhao is with the Department of Radiation Oncology, Stanford University, Palo Alto, CA 94306, USA (e-mail: zhaow85@stanford.edu).}}
\maketitle

\begin{abstract}
    Dual-energy computed tomography (DECT) is of great significance for clinical practice due to its huge potential to provide material-specific information. However, DECT scanners are usually more expensive than standard single-energy CT (SECT) scanners and thus are less accessible to undeveloped regions. In this paper, we show that the energy-domain correlation and anatomical consistency between standard DECT images can be harnessed by a deep learning model to provide high-performance DECT imaging from fully-sampled low-energy data together with single-view high-energy data, which can be obtained by using a scout-view high-energy image. We demonstrate the feasibility of the approach with contrast-enhanced DECT scans from 5,753 slices of images of twenty-two patients and show its superior performance on DECT applications. The deep learning-based approach could be useful to further significantly reduce the radiation dose of current premium DECT scanners and has the potential to simplify the hardware of DECT imaging systems and to enable DECT imaging using standard SECT scanners.
\end{abstract}


\section{Introduction}
\label{sect:intro}  % \label{} allows reference to this section
Material differentiation and quantification using a standard single-energy computed tomography (SECT) is extremely challenging because different materials may have the same CT value\cite{ref1}. To tackle this challenge, dual-energy CT (DECT) takes full advantage of the energy dependence of the linear attenuation coefficient by scanning the patients using two different energy spectra\cite{ref2, ref3, ref4, ref5, ref6,lee2017feasibility,petrongolo2018single,xue2019accurate}. This enables DECT imaging providing energy- and material-selective images, and having been very widely used in clinical practice for many applications, such as virtual monochromatic imaging\cite{ref7, ref8}, differentiating intracerebral hemorrhage from iodinated contrast\cite{ref9}, automated bone removal in CT angiography\cite{ref10, ref11, ref12, ref13}, virtual noncontrast-enhanced imaging\cite{ref14, ref15, ref16, ref17, ref18, ref19} and urinary stone characterization\cite{ref20, ref21, ref22, ref25}. However, it is still an open and challenging task for clinical DECT imaging due to complex practical implementations, proprietary patents for major CT vendors, and less popularity for DECT scanners compared to the standard SECT scanners.

Since the low- and high-energy CT images acquired from the DECT scanners have the same anatomical structures, there is substantial redundant anatomical information between the DECT images. For the scanned patients using the same DECT imaging protocols, the low- and high-energy CT images are also correlated in the energy domain, resulting in information redundancies in the energy-domain\cite{ref26, ref27}. Meanwhile, both DECT images are reconstructed using fully-sampled projection data which have to meet the classical Shannon-Nyquist theorem in angular-data sampling to reconstruct artifacts-free images. By fully exploiting the anatomical consistency and energy domain correlation between the DECT images, it is possible to provide high-quality artifacts-free DECT images using conventional SECT images together with sparse sampling projection data at different energy levels.

\begin{figure*}[t]%[th!b]
    \centering
        \includegraphics[width=0.8\textwidth]{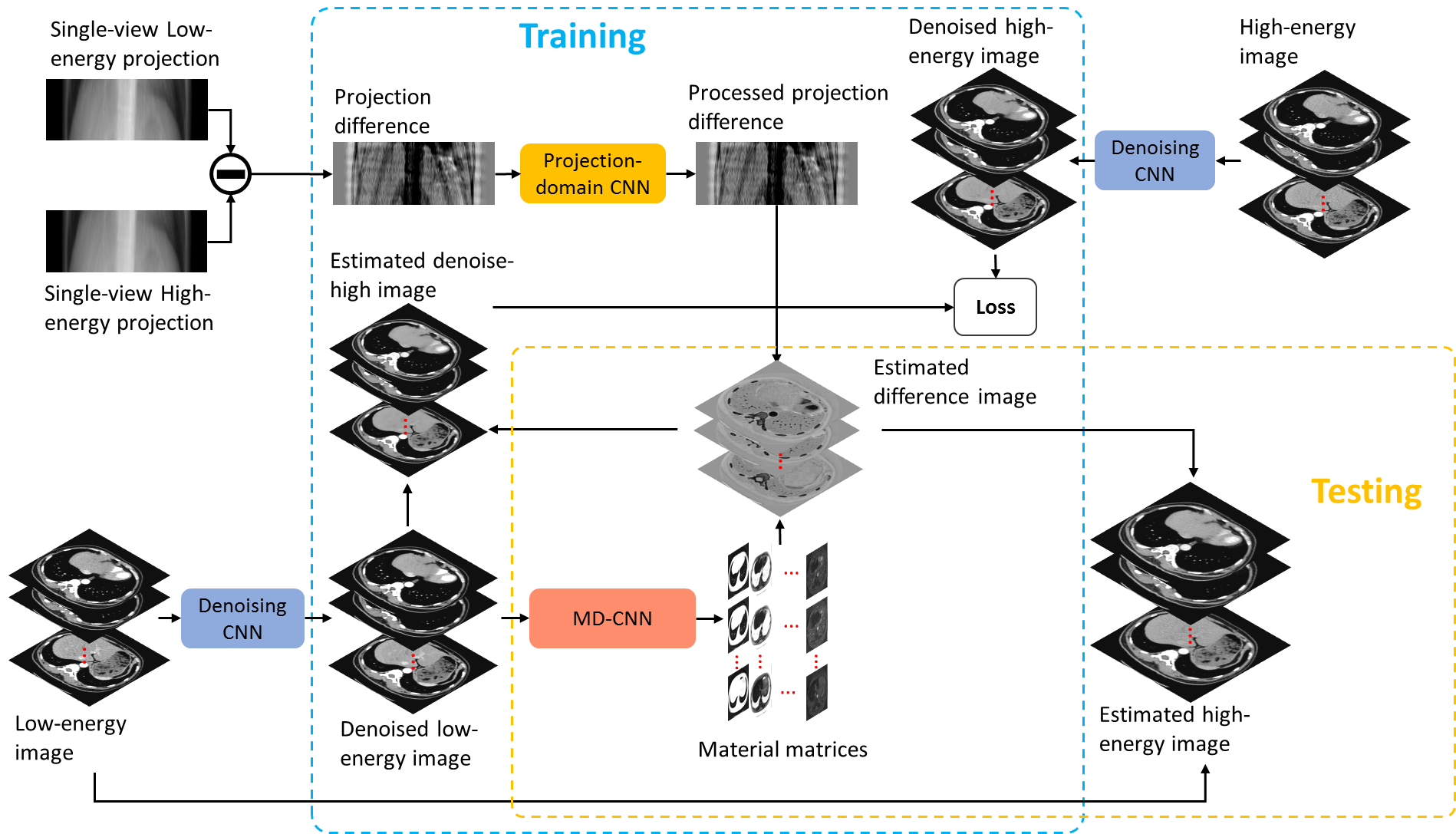}
        \caption{The workflow of the proposed fully-sampled low-energy and single-view high-energy DECT imaging approach. During the training phase, the denoising DECT images together with the single-view dual-energy projections are used to train the projection domain convolutional neural network (CNN) and the MD-CNN. In the testing phase, the trained networks use the input low-energy images and the single-view dual-energy projections to infer the corresponding high-energy images.}
    \label{fig:1}
\end{figure*}

CT imaging with the full use of as low as reasonably achievable (ALARA) principle has been commonly accepted in routine practice and further reducing radiation dose from CT scanning is clinically favorable and has been extensively studied for almost two decades. Deep learning (DL) has recently been proved to be a powerful tool for mapping complex relationships and incorporating existing knowledge into an inference model through feature extraction and representation learning\cite{ref28, ref29, ref30, ref31, ref32, ref33, ref34, ref35, ref36, ref37}. It has also been applied in low-dose CT\cite{ref38, ref39, ref40, ref41} and DECT imaging\cite{ref42, ref43, ref44, ref45, ref46}.

To further significantly reduce the radiation dose of DECT imaging, in this study, we synergically exploit the energy-domain correlation and anatomical consistency between DECT images by leveraging the deep learning approach and the seamless integration of the correlation and consistency in a data-driven DECT imaging process, and eventually push the sparse sampling to the limit of a single projection view and demonstrate the feasibility of high-performance DECT imaging using a deep learning approach termed fully low-energy and single high-energy DECT imaging (FLESH-DECT).

%In this study, we push the sparse sampling to the limit of a single projection view and demonstrate DECT imaging using the standard SECT imaging together with a single-view of the second energy level. To achieve this, we take advantage of the energy-domain correlation and anatomical consistency between the low- and high-energy images by leveraging deep learning approach and the seamless integration of the correlation and consistency in a data-driven DECT imaging process.

\section{Method}
\label{sect:method}
The flowchart of proposed FLESH-DECT strategy is shown in Fig.~\ref{fig:1}.
The input to the model is a single-view high-energy projection together with the low-energy image \(I_{low}\) which is reconstructed using fully-sampled low-energy projection data. For the low-energy image, it is firstly denoised with a denoising network to mitigate the impact of image noise. Instead of training a network directly mapping high-energy images from the low-energy images, we use a convolutional neural network (CNN) to perform material-decomposition-type operations and it is termed material decomposition CNN (MD-CNN). The input of the MD-CNN is the denoised low-energy image \(I_{low}^{de}\) while the output is a "material component" matrix \(A\). The matrix \(A\) has the same image size as \(I_{low}^{de}\) but with multiple channels each of which corresponds to a pseudo material-specific image. The values are the percentages of corresponding "basis material" on these pixels. We ensure that the sum of the percentages equals to one (mass conversation) for each unique pixel. Furthermore, another CNN is used to pre-process the differences between the given high-energy projections and its corresponding low-energy projections. This projection-domain network
is used to fill the gap between the denoised images and the non-denoised projection. Since CT forward projection can be regarded as linear summations of pixel values, we can use the least squares method to solve the corresponding CT values of each "material" \(b_{dif}\) according to the matrix \(A\) and the pre-processed projection difference. The estimated high-energy image is calculated as the summation of low-energy images, and the inner product of matrix \(A\) and vector \(b_{dif}\).
Detailed formula derivation is described in the following subsections.

\subsection{DECT Imaging}
In CT imaging, the attenuation coefficient at each position can be represented as a linear combination of basis materials’ attenuation coefficient\cite{ref50}.
\begin{equation}
    \mu=\alpha_1\mu_1+\alpha_2\mu_2+\dots+\alpha_m\mu_m
    \label{eq:1}
\end{equation}
where \(m\) is the number of basis materials, \(\alpha_i\) is the percentage of the \(i\)-th basis material and \(\mu_i\) is the attenuation coefficient of the \(i\)-th basis material. Since CT values in Hounsfield Unit (HU) can be represented as the linear transformation of attenuation coefficient \(\mu\) with the following equation:
\begin{equation}
    HU=1000\times\frac{\mu-\mu_{water}}{\mu_{water}}
    \label{eq:2}
\end{equation}
Eq.\eqref{eq:1} can also be written as:
\begin{equation}
    HU=\alpha_{1}HU_{1}+\alpha_{2}HU_{2}+\dots+\alpha_{m}HU_{m}
    \label{eq:3}
\end{equation}
where \(HU_{i}\) stands for the Hounsfield Unit CT value for the \(i\)-th basis material.
Considering there are \(n_{pix}=W{\times}H\) pixels in an image slice, we can write Eq.\eqref{eq:3} into the following matrix multiplication form
\begin{equation}
    I=A\cdot{b}^T
    \label{eq:4}
\end{equation}
where \(I\) is the image vector sized \(n_{pix}{\times}1\) containing  CT values at each pixel, \(A={[\alpha_{ij}]}_{n_{pix}{\times}m}\) is the material component matrix, \(\alpha_{ij}\) stands for the percentage of material \(j\) at pixel \(i\), \(b={[HU_i]}_{m{\times}1}\) consists of HU values for each material.

In DECT, there are two different images \(I_{low}\) and \(I_{high}\). The material component matrix \(A\) remains the same for both images because pixel compositions do not change between low- and high-energy scans. Therefore, we have the following equations
\begin{equation}
    \begin{cases}
        &I_{low}=A\cdot{b}_{low}^T\\
        &I_{high}=A\cdot{b}_{high}^T
    \end{cases}
    \label{eq:5}
\end{equation}
By subtracting the high-energy equation from the low-energy equation, we get
\begin{equation}
    I_{dif}=A\cdot{b}_{dif}^T
    \label{eq:6}
\end{equation}
where \(I_{dif}\) is the difference image between \(I_{low}\) and \(I_{high}\), \(b_{dif}\) is the difference between \(b_{low}\) and \(b_{high}\). Let \(P_{high}\) and \(P_{low}\) be the given high- and low- energy projection measurements, and \(R\) be the projection matrix sized \(n_{ray}{\times}n_{pix}\) corresponding to the high-energy view. We have the following equation
\begin{equation}
\begin{split}
    R{\cdot}A\cdot{b}_{dif}^T & = R{\cdot}I_{dif} \\
                               & = R{\cdot}I_{high}-R{\cdot}I_{low} \\
                               & = P_{high}-P_{low}
    \label{eq:7}
\end{split}
\end{equation}
For the fully-sampled low-energy and single-view high-energy CT imaging task, the unknows in Eq.\eqref{eq:7} are the material component matrix \(A\) and the corresponding difference values \(b_{dif}\). Assuming that we have the material component matrix \(A\), let \(M=R{\cdot}A\in\Re^{m{\times}n_{ray}}\), \(P_{dif}=P_{high}-P_{low}\in\Re^{1{\times}n_{ray}}\), the difference values \(b_{dif}\) can be calculated by solving the equation \(M{\cdot}b^{T}=P_{dif}\). In regular CT imaging, we have \(n_{ray}>>m\), the best \(b_{dif}\) can therefore be found using least-squares method which can be computed with Cholesky decomposition, i.e.,
\begin{equation}
\begin{split}
    b_{dif} & =\operatornamewithlimits{argmin}\limits_{\tilde{b}}{\|M{\cdot}\tilde{b}^T-P_{dif}\|}_2^2 \\
            & =[(M^TM)^{-1}M^TP_{dif}]^T
    \label{eq:8}
\end{split}
\end{equation}
The only task now is to estimate the material component matrix \(A\) from the low energy image \(I_{low}\).

\subsection{Material decomposition-based dual-energy CT mapping}
Due to its ability to learn complex relationships and incorporate existing knowledge into a nonlinear mapping model, a dedicated CNN model (termed MD-CNN) is used to estimate the material component matrix \(A\). When designing the MD-CNN model, a major challenge is the lack of training labels due to unknown materials in the images and their percentages. To tackle this challenge, we train the MD-CNN indirectly. We firstly denoise the dual-energy image pairs, and the denoised low-energy images \(I_{low}^{de}\) are inputted into the MD-CNN to acquire material component matrices \(A_{DL}\). Meanwhile, we put the projection differences into another 1-D projection domain CNN to preprocess the projection data. Then, we compute \(b_{dif}\) using Eq.\eqref{eq:8}. The estimated denoised high-energy images \(I_{high}^{dl}\) can therefore be calculated as
\begin{equation}
    I_{high}^{dl}=I_{low}^{de} + A_{DL}{\cdot}b_{dif}^T
    \label{eq:10}
\end{equation}
An image similarity loss is calculated between the denoised high-energy image \(I_{high}^{de}\) and the DL-estimated image \(I_{DL}^{de}\), and the mean squared error (MSE) loss is used for the task:
\begin{equation}
    \mathcal{L}_{high}=\frac{1}{n}{\|I_{high}^{de}-I_{DL}^{de}\|}_2^2
    \label{eq:11}
\end{equation}
Instead of getting material component label, we focus on the target high-energy image and it is not necessary to specify each channel in \(A_{DL}\) to represent real material or linear combination of different materials. Meanwhile, since matrix \(A_{DL}\) is supposed to be the material component matrix, it should be able to recover the input denoised low-energy image as well, resulting in the follow loss function:
\begin{equation}
    \mathcal{L}_{low}=\frac{1}{n}{\|I_{low}^{de}-A_{DL}{\cdot}b_{low}^T\|}_2^2
    \label{eq:12}
\end{equation}
The same strategy in Eq.\eqref{eq:8} is used to calculate the HU values \(b_{low}\) for each "material" under low energy,
\begin{equation}
    b_{low}=\operatornamewithlimits{argmin}\limits_{\tilde{b}}{\|A_{DL}{\cdot}\tilde{b}^T-I_{low}^{de}\|}_2^2
    \label{eq:13}
\end{equation}
The final loss function is computed as the summation of \(\mathcal{L}_{low}\) and \(\mathcal{L}_{high}\), i.e.,
\begin{equation}
    \mathcal{L}=\mathcal{L}_{low}+\mathcal{L}_{high}
    \label{eq:14}
\end{equation}

During the inference phase, the low-energy images are also firstly denoised and then inputted into the trained MD-CNN for \(A_{DL}\). The projection domain CNN preprocesses the projection difference \(P_{dif}\). The estimated difference images are calculated according to Eq.\eqref{eq:8}. The difference between the training and inference is the estimated high-energy image is calculated as the summation of the original low-energy image \(I_{low}\) and the difference image \(I_{dif}\) in inference phase.

\subsection{Network details}
\subsubsection{Denoising CNN}

\begin{figure}[t]%[th!b]
    \centering
        \includegraphics[width=0.48\textwidth]{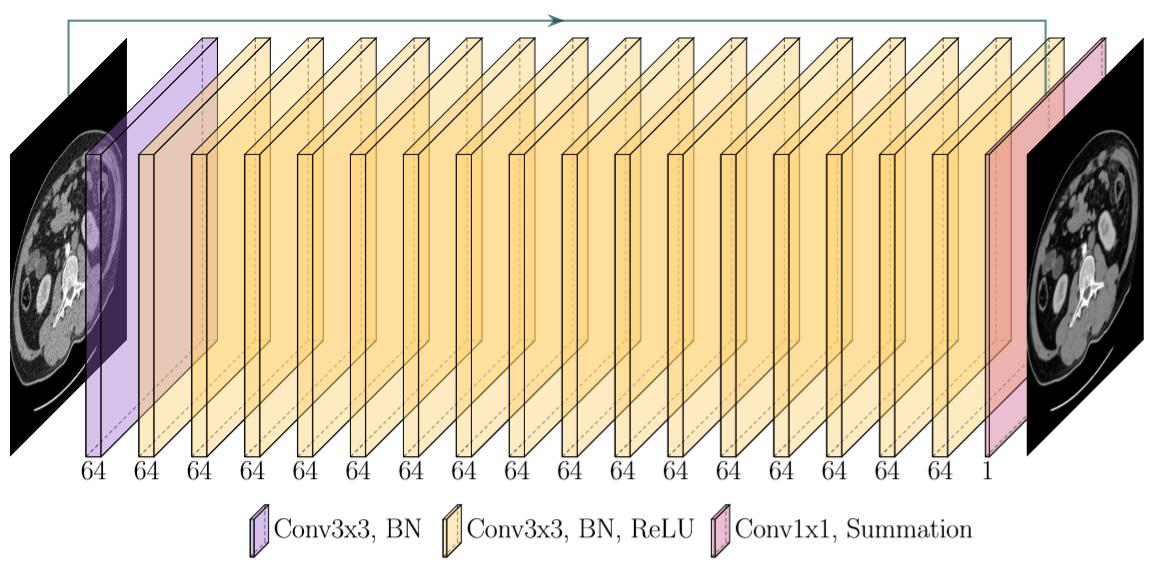}
        \caption{Architecture of the fully convolutional network for image denoising. A plain structure which encompasses 13 convolution layers is applied to learn the residual between the input image and the denoised image. }
    \label{fig:2}
\end{figure}

We employ the denoising network in our previous work\cite{ref46} to reduce the DECT image noise. The network uses a plain structure which encompasses 13 convolution layers to learn the residual between the input image and the denoised image. The first 12 layers are convolution layers with kernel size \(3\times3\) and each layer is followed by a batch normalization layer (BN) and a rectified linear unit (ReLU) activation. The last layer is a convolution layer with kernel size \(1\times1\) fusing the result. Fig.~\ref{fig:2} shows the detailed structure of the denoising CNN. The denoised image is computed as the summation of the input image and the output from the last layer.

\subsubsection{MD-CNN}

\begin{figure}[t]%[th!b]
    \centering
        \includegraphics[width=0.48\textwidth]{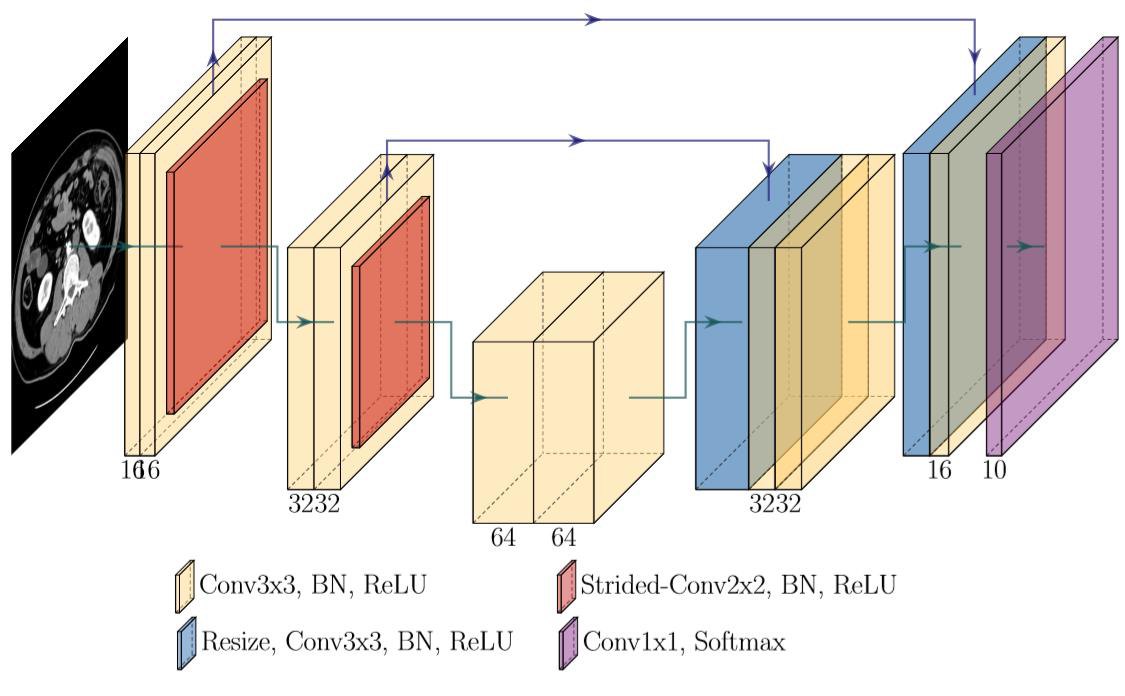}
        \caption{Structure of the MD-CNN. A much simplified UNet-like structure with 14 convolution layers is used here to estimate the percentages of each corresponding "basis material". Numbers under each block show
the number of channels of the multichannel feature maps.}
    \label{fig:3}
\end{figure}

For the material decomposition network, we employ a U-Net-type structure\cite{ref51} which has a large receptive field and is quite suitable for many medical image processing tasks. There are 10 normal \(3\times3\) convolution layers in the proposed network (Fig.~\ref{fig:3}). Each convolution layer is followed by a BN layer and a ReLU activation layer. There are 3 resolution levels in total. For down-sampling, we use convolution layer with kernel size
of \(2\times2\) and stride equal to 2. Each strided convolution layer is also followed by a BN layer and a ReLU activation layer. For up-sampling, we use bilinear interpolation to double both image width and height. At the end of the network, a convolution layer with kernel size
of \(1\times1\) is added to fuse the channels. Since the values on each output channel are supposed to be the percentages of corresponding "basis material", we apply softmax after the convolution layer with kernel size
of \(1\times1\) to make sure that the sum of materials’ proportion equals to 1 at each pixel.

\subsubsection{Projection-domain CNN}

\begin{figure}[t]%[th!b]
    \centering
        \includegraphics[width=0.35\textwidth]{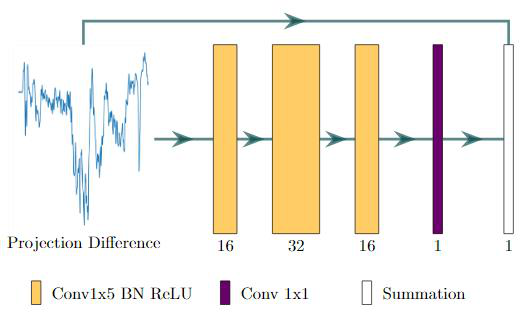}
        \caption{Structure of the projection-domain CNN. A concise 4-layer network is employed to slightly refine the inputting projection difference. Numbers under each block show the number of channels of the multichannel feature maps.}
    \label{fig:4}
\end{figure}

To enhance the robustness of the lease-square problem in Eq.\eqref{eq:8}, the projection-domain network (Fig.~\ref{fig:4}) is applied to slightly refine the inputting projection difference and to make its noise level to be consistent with that of the denoised DECT difference image, and we employ a concise 4-layer network for the task. A residual learning structure is used here for an easier startup at the beginning iterations. All basic convolution layers have a kernel with size of \(1\times5\) except for the last one which has a kernel with size of \(1\times1\). The first three convolution layers are followed by a BN layer and a ReLU activation layer.

\subsubsection{Network training}
The denoising CNN was implemented in MATLAB with MatConvNet framework\cite{ref47}. Denoising was performed as a data preprocessing step for all DECT images. The material decomposition network and projection-domain network are implemented using Python with Tensorflow framework\cite{ref48}. Those two networks were trained together in an end-to-end fashion. The parameters in the networks were optimized using ADAM algorithm\cite{ref49} with \(\beta_1=0.9\) and \(\beta_2=0.999\). Learning rate was set to \(10^{-3}\) during the training. The training set was randomly split into small batches in each epoch with batch size of 8. The proposed network was trained for 200 epochs in total. We perform validation after each training epoch and the model with best validation loss was selected as the final model for testing. The number of materials \(m\) was set to 10 in our experiments. It has to note that the model was able to achieve reasonable results even with \(m=3\). The networks were trained and tested on a workstation with configurations as follows: CPU is Intel(R) Xeon(R) Gold 6130 CPU @ 2.10GHz; GPU is NVIDIA RTX 2080 Ti with 12G memory.

\subsection{Projectors for different CT geometries}
In order to calculate matrix \(M\) in Eq.\eqref{eq:8}, projector corresponding to the CT geometry is indispensable in our algorithm. There are mainly two types of 2D CT geometries, fan-beam and parallel-beam. The fan-beam geometry can be further divided into two sub-types, equiangular and equispacing. For each above-mentioned geometry, we developed a projector and trained a new model for evaluation.
\subsubsection{Equiangular fan-beam}
\begin{figure}[tb]
    \centering
        \includegraphics[width=0.48\textwidth]{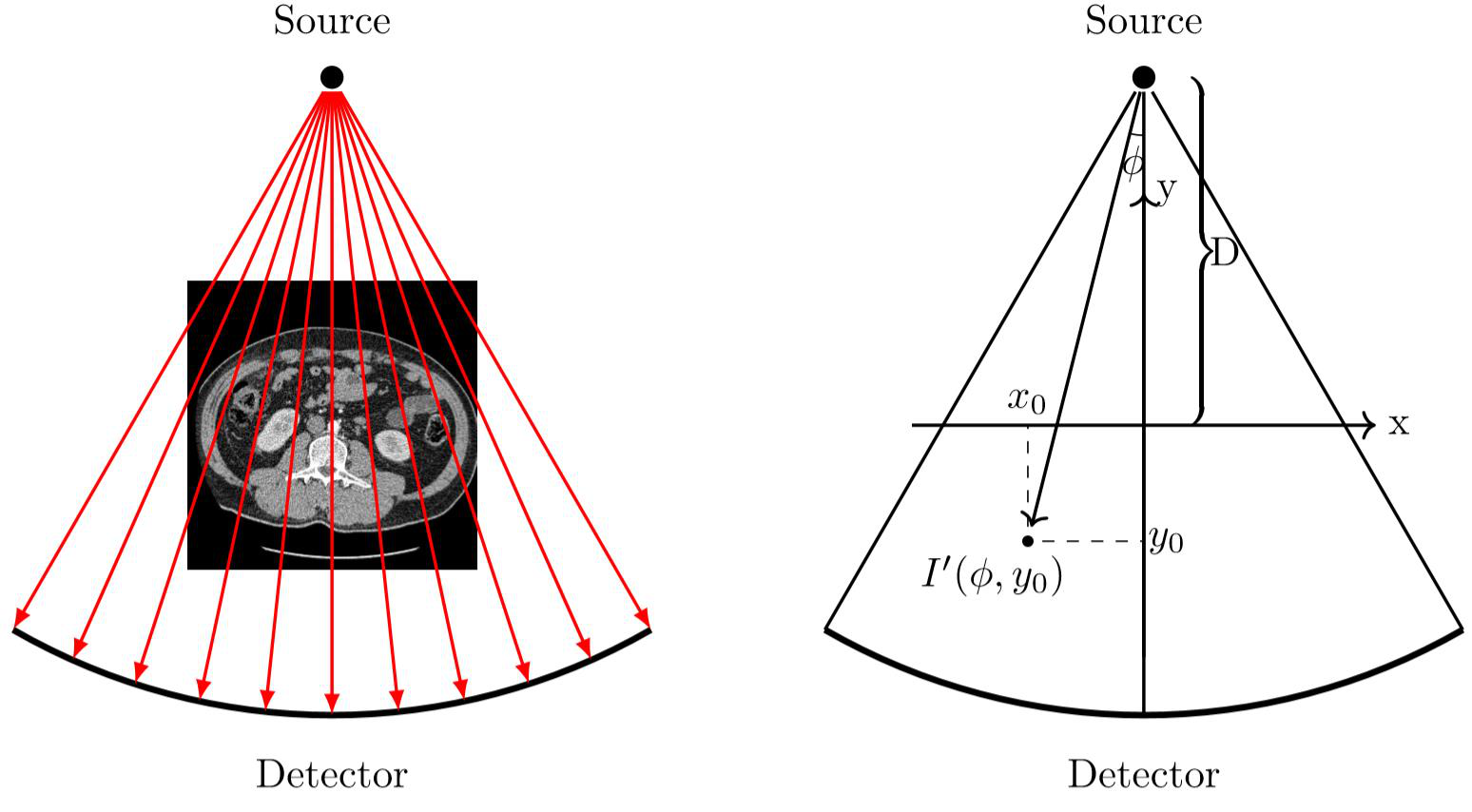}
        \caption{Illustration of equiangular fan-beam geometry. }
    \label{fig:8}
\end{figure}

Equiangular geometry is mainly implemented with arc detectors which keep the angles between two adjacent detector pixels and the source-detector-distance the same. We test our model in the anterior-posterior (AP) direction, but it can be easily extended to any other projection view.
%A simple implementation of equiangular fan-beam projection is adopted here for dataset preparing and the projection in the model.
To acquire projection data using the 2D equiangular fan-beam geometry, we first calculate the intersection of each projection ray with each image row. The values at each intersection points are computed using linear interpolation. Suppose the image size is \(W{\times}H\), and a matrix \(I'\) with size of \(n_{ray}{\times}H\) can be obtained using the following rebinning equation:
\begin{equation}
    I'(\phi, y)=I((D-y)\tan(\phi), y)
    \label{eq:15}
\end{equation}
where \(\phi\) is the angle between the projection ray and the central ray (as shown in Fig.~\ref{fig:8}), and \(y\) is the image pixel position along y-axis in the Cartesian coordinate system centered at image center \(O\). \(D\) is the distance between projection source and the rotation center which is overlapped with the image center \(O\). After the linear interpolation, we calculate the summation of each column in matrix \(I'\) to obtain raysum $S$ which is weighted by the distance to yield the final projection $P$:
\begin{equation}
    P(\phi)=\frac{dy}{\cos{\phi}}S(\phi)
    \label{eq:16}
\end{equation}
where \(dy\) is the image spacing in y-axis.
%Fig. \ref{fig:8} depicts the sketch map of equiangular fan-beam projection and the definition of the parameters.
We set \(D=600mm\), \(dy=0.5mm\), the number of detector channels \(n_{ray}=800\) and the angle between adjacent channels \(ds=\frac{7}{11000}rad\) for all images.

\subsubsection{Equispacing fan-beam}
\begin{figure}[tb]
    \centering
        \includegraphics[width=0.48\textwidth]{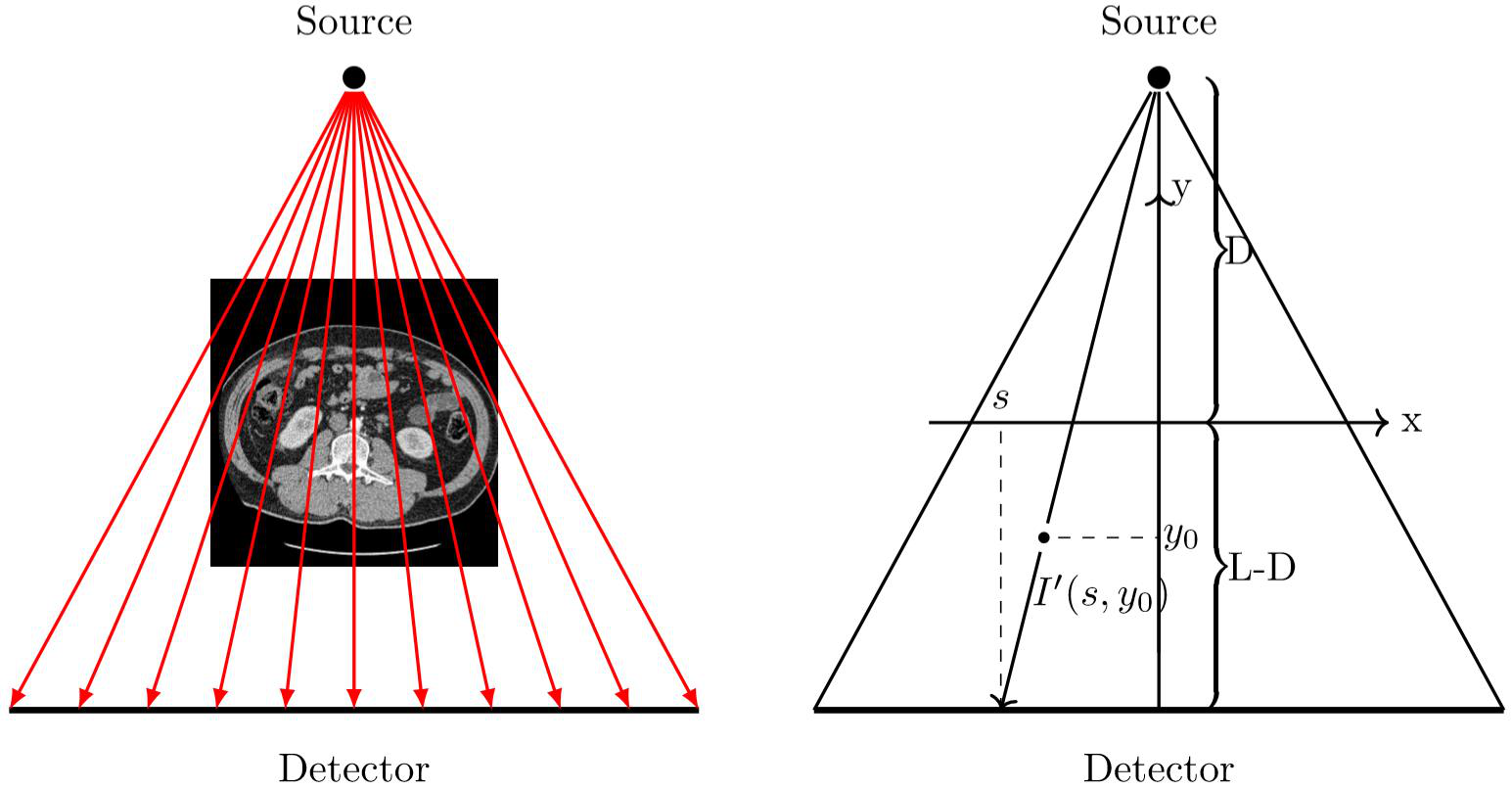}
        \caption{Illustration of equispacing fan-beam geometry. }
    \label{fig:9}
\end{figure}

Flat detectors with equal spacing between the adjacent channels are commonly used to implement the equispacing geometry. We use similar strategy to implement the equispacing fan-beam projection in the AP direction.
In this case, the rebinning procedure is computed with the following equation:
\begin{equation}
    I'(s, y)=I(\frac{s(D-y)}{L}, y)
    \label{eq:17}
\end{equation}
where \(s\) is the distance between current channel and the detector center (direction included), and \(L\) is the source to detector distance, as shown in Fig.~\ref{fig:9}. The rest part of the implementation is the same as equiangular projection. For the parameters, we have \(D=600mm\), \(L=1100mm\), \(dy=0.5mm\), \(n_{ray}=800\) and the spacing between nearby channels \(ds=0.78mm\). %The equispacing geometry is illustrated in .

\subsubsection{Parallel-beam}
%Since we are calculating parallel-beam projection in the AP direction, a very simple implementation is adopted here.
For the parallel-beam scenario, We assume that the detector channel has the same spacing as the image pixels and each X-ray projects exactly through an image column in the AP direction. Therefore, the projection in the AP direction can be computed as the summation of each image column.

\section{Data Specification}
\subsection{Training data for the denoising network}
The AAPM Low-Dose CT Grand Challenge data are used to train the denoising network. This dataset consists of routine dose CT and the corresponding simulated low-dose CT data from 10 patients. The routine dose scanning voltage is 100 kV or 120 kV and the X-ray tube current varies from 200mA to 500mA. The detector has \(736\times64\) elements, and each element has a size of \(1.2856\times1.0947mm^2\). The source-to-axial distance is \(59.5cm\) and the source-to-detector distance was \(108.56cm\). All the images were reconstructed to slice thickness of \(1.0 mm\) and \(512\times512\) pixel. The pixel size varies from \(0.66\times0.66mm^2\) to \(0.78\times0.78mm^2\). To simulate the low-dose CT data, Poisson noise was introduced into the routine dose to mimic a noise level that corresponded to \(25\%\) of the routine dose, and noisy projection are reconstructed to yield the low-dose image.  %\cite{ref52}

\subsection{DECT imaging dataset}
Clinical DECT images of 22 patients who underwent iodine contrast-enhanced DECT exams were collected for the study. All the exams were performed in Nanjing General PLA Hospital, China, with the approval by the institutional review board and patient consent forms. The DECT images (5753 slices in total) were acquired using a SOMATOM Definition Flash DECT scanner (Siemens Healthineers, Forchheim, Germany) after administering iodine contrast agent. The low- and high-energy of the DECT scans were 100 kV and 140 kV, respectively. All CT images were reconstructed using the filtered back-projection (FBP) algorithm provided by the commercial CT vendor. The dataset were split into training set, validation set and testing set randomly with 16, 3 and 3 patients included respectively.

\begin{figure}[t]%[th!b]
    \centering
        \includegraphics[width=\linewidth]{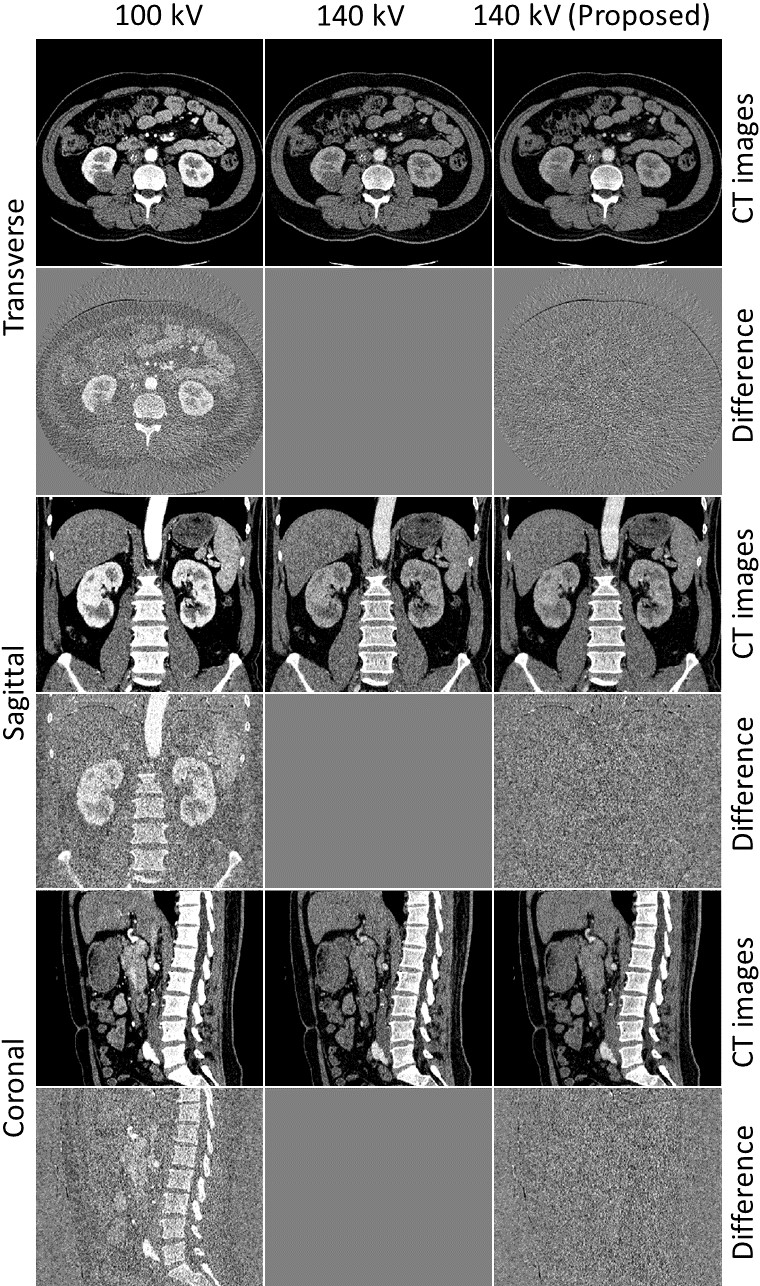}
        \caption{Example results on a testing slice and the difference with respect to the corresponding real high-energy image. The first, third and fifth row display the images on axial, sagittal and coronal view. From left to right are real 100 kV image, real 140 kV image and result from proposed method, respectively. The second, fourth and sixth row are the corresponding differences with real 140 kV images. The CT images are displayed with a window width=300 HU and center=50 HU while the difference images are displayed under window width=300 HU and center=0 HU.}
    \label{fig:5}
\end{figure}

\section{Results}
%The single-view high-energy projection for the FLESH-DECT method was acquired in the AP direction using equiangular fan-beam CT geometry. Projection acquired using other CT geometries (such as equispacing fan-beam geometry) was also investigated and shown similar results (details in Supplementary materials).

We first focus on the results of equiangular geometry, and comparison results using different geometries are presented at the end this section.
Fig.~\ref{fig:5} shows original DECT images and the 140 kV images predicted using the proposed method for a testing patient. The first, second and third columns show the original 100 kV images, the original 140 kV images, and the predicted results, respectively. The first, third, and fifth rows show CT images in transverse, sagittal, and coronal planes, respectively. The second, fourth, and sixth rows show difference images with respect to the corresponding real high-energy CT images in transverse, sagittal, and coronal planes, respectively.
\begin{table}[t]%[th!b]
    \begin{center}
        \caption{Quantitative comparisons between the predicted and the real high-energy CT images for the testing patients. }
        \begin{tabular}{ c | c c c c c }
            \hline\hline
             & MSE & PSNR & SSIM & HU error & Time(s) \\
            \hline
            Patient1 & 875.06 & 36.80 & 0.8702 & 1.65\(\pm\)1.03 & 2.65 \\
            Patient2 & 887.60 & 36.99 & 0.8789 & 2.09\(\pm\)1.48 & 2.31 \\
            Patient3 & 927.47 & 36.89 & 0.8745 & 1.50\(\pm\)0.82 & 3.34 \\
            \hline\hline
        \end{tabular}
        \label{tab:1}
        %\vspace{-5mm}
    \end{center}
\end{table}
\begin{figure}[t]%[th!b]
    \centering
        \includegraphics[width=\linewidth]{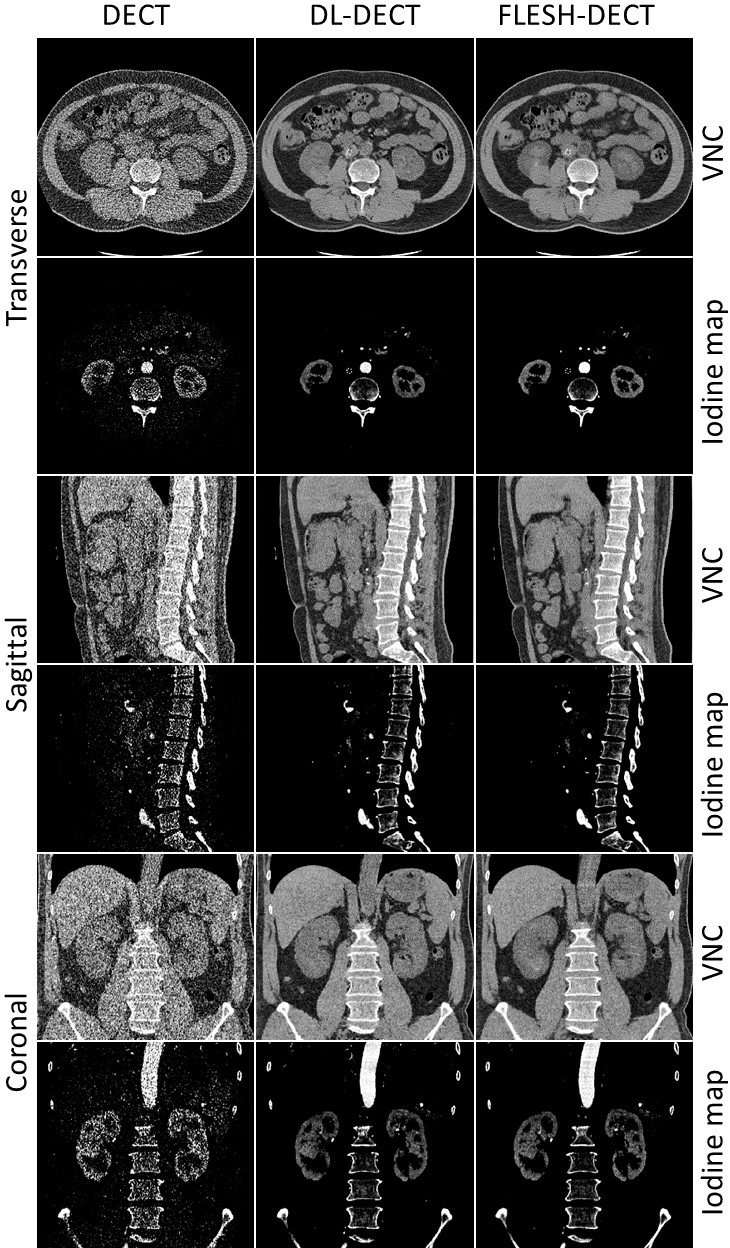}
        \caption{VNC images and Iodine maps reconstructed using original DECT, DL-DECT, and FLESH-DECT images. The first, third and fifth row are the VNC images on axial, sagittal and coronal view, respectively, while the second, fourth and sixth row are the corresponding Iodine maps. }
    \label{fig:6}
\end{figure}
As can be seen, the proposed DL-derived high-energy images are highly consistent with the original high-energy images. There are some differences at sharp boundaries which also appear in difference images between original high- and low-images. Those differences may be motion introduced difference between original DECT images because there is approximately 90 degrees out of phase for the low- and high-energy data acquired using a dual-source DECT scanner. When inputting the low-energy image into the model, the model performed prediction based on the anatomical structure of the low image and can not reflect the change with respect to the original high-energy image.

%\vspace{-5mm}
\begin{table}[b]%[th!b]
    \begin{center}
        \caption{Quantitative comparisons of the VNC image and the iodine maps reconstructed using original DECT and FLESH-DECT images.}
        \begin{tabular}{ c c | c c | c c }
            \hline\hline
            \multicolumn{2}{c}{\multirow{2}{*}{}} & \multicolumn{2}{|c|}{VNC} & \multicolumn{2}{c}{Iodine Map} \\
            & & MSE & PSNR & MSE & PSNR \\
            \hline
            \multirow{2}{*}{Patient1} & DL-DECT & 3882.34 & 29.03 & 0.0313 & 23.65 \\
            & FLESH-DECT & \textbf{3808.97} & \textbf{29.11} & \textbf{0.0308} & \textbf{23.73} \\
            \hline
            \multirow{2}{*}{Patient2} & DL-DECT & 4216.36 & 28.95 & 0.0342 & 23.78 \\
            & FLESH-DECT & \textbf{4128.01} & \textbf{29.04} & \textbf{0.0335} & \textbf{23.87} \\
            \hline
            \multirow{2}{*}{Patient3} & DL-DECT & 3508.35 & 29.71 & 0.0313 & 24.85 \\
            & FLESH-DECT & \textbf{3431.13} & \textbf{29.81} & \textbf{0.0306} & \textbf{24.95} \\
            \hline\hline
        \end{tabular}
        \label{tab:2}
    \end{center}
\end{table}

Quantitative metrics were calculated to evaluate the accuracy of the predicted high-energy CT images. We use the well-established metrics MSE, PSNR and SSIM to assess the image similarity between the real and the predicted high-energy images. Additionally, more than 100 region-of-interests (ROIs) were randomly selected for each testing volume on homogeneous areas (e.g. liver and stomach). We calculated the mean HU value differences on those ROIs and the results show that the averaged HU error between the predicted and the original high-energy image is smaller than 2.09 HU. For the computation time, it takes around 2.5 seconds for the proposed method to process 300 slices. All quantitative results are shown in Table \ref{tab:1}.
Since most of the differences come from noise, we also compared the denoised DL-predicted images with the denoised high-energy images. In this case, the DL-predicted images are calculated by adding the denoised low-energy image to the DL-estimated difference image. For those denoised images, the proposed method achieves average MSE of 171.09, PSNR of 44.33 and SSIM of 0.9848.

In our previous work \cite{ref46}, we propose the DL-DECT method to estimate \(I_{dif}\) directly from \(I_{low}\). In this work, with the introduction of the additional high-energy single-view projection, we can further enhance the accuracy of the predicted high-energy image and eventually the accuracy of the material- and energy-specific images. To show the benefit of the additional high-energy projection, we quantitatively compared the proposed FLESH-DECT method with the DL-DECT method.
Virtual noncontrast (VNC) images and iodine maps are derived to demonstrate the clinical utility of the proposed method. Fig. \ref{fig:6} depicts the VNC images and iodine maps reconstructed using different methods on the transverse, sagittal and coronal planes. Both the DL-DECT and FLESH-DECT algorithms provide high-quality VNC and iodine images that are consistent with the images generated by original DECT images. Quantitative metrics on VNC and iodine images are shown in Table \ref{tab:2}. The results demonstrate FLESH-DECT can provide high-quality material-specific images and it outperforms the DL-DECT method.

\begin{table}[t]%[th!b]
    \begin{center}
        \caption{Quantitative Noise Level Comparison on VNCs and Iodine Maps.}
        \begin{tabular}{ c c | c c }
            \hline\hline
            \multicolumn{2}{c}{Standard Deviation on ROIs} & VNC & Iodine Map \\
            \hline
            \multirow{2}{*}{Patient1} & Real & 81.58 & 0.3363 \\
            & DL-DECT & 29.14 & 0.1618 \\
            & FLESH-DECT & \textbf{28.09} & \textbf{0.1499} \\
            \hline
            \multirow{3}{*}{Patient2} & Real & 74.69 & 0.2837 \\
            & DL-DECT & 24.81 & 0.1017 \\
            & FLESH-DECT & \textbf{24.12} & \textbf{0.0998} \\
            \hline
            \multirow{3}{*}{Patient3} & Real & 76.95 & 0.2875 \\
            & DL-DECT & 25.97 & 0.0946 \\
            & FLESH-DECT & \textbf{25.67} & \textbf{0.0891}\\
            \hline\hline
        \end{tabular}
        \label{tab:3}
        %\vspace{-5mm}
    \end{center}
\end{table}

From the VNC images and Iodine maps, we find that the DL-derived VNC images and iodine maps show a remarkably reduced noise level compared with those generated from original DECT images. Here we also compare the noise level by calculating the standard deviation in ROIs. More than 500 ROIs were selected randomly in homogeneous areas. The mean standard deviations in ROIs on each testing patient are provided and compared in Table \ref{tab:3}. The mean standard deviations of the images derived using FLESH-DECT are close to that of the DL-DECT method which is much lower than those from the original DECT images. This result shows that the FLESH-DECT method maintains the denoising feature.

\begin{figure*}[tb]
    \centering
        \includegraphics[width=0.98\textwidth]{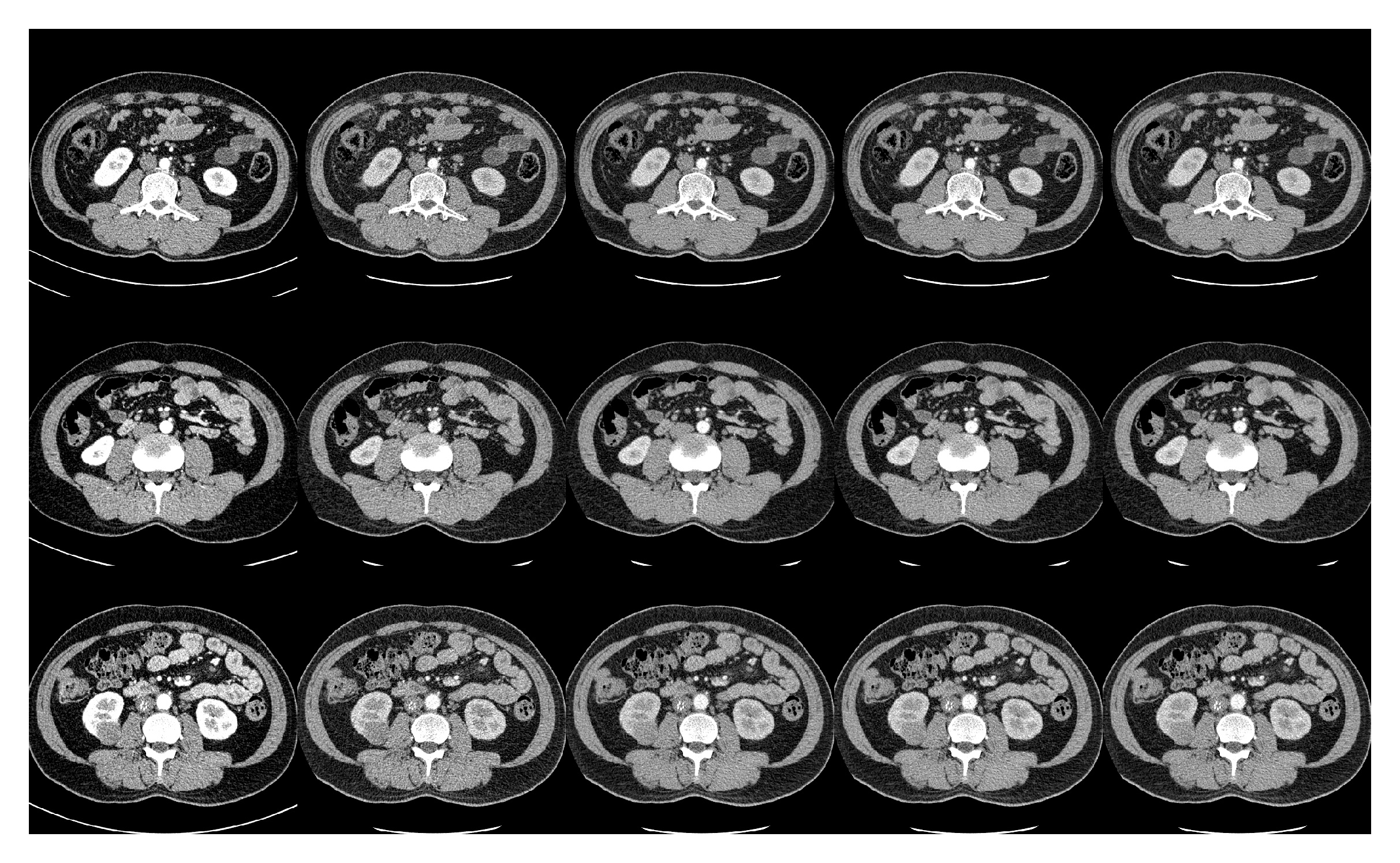}
        \caption{Results on three testing slices for different geometry models. From left to right are original 100 kV images, original 140 kV images, parallel-beam results, equispacing fan-beam results and equiangular fan-beam results, respectively. }
    \label{fig:7}
\end{figure*}

We tested the proposed method with several 2D CT geometries. Example results on a testing volume using different geometries are displayed in Fig. \ref{fig:7}. All CT images are displayed with window width=300HU and center=50HU while difference images are displayed with window width=300HU and center=0HU. As can be seen, all models are able to provide competitive results which are highly consistent with the original high energy image. For better comparison, we also calculate quantitative metrics on those results (Table \ref{tab:4}). The differences between the results obtained using the proposed method with different geometries are marginal and all of them are superior to our previous DL-DECT method which does not utilize the additional single-view projection. Overall, the proposed method reduces mean-squared error (averaged for all testing cases) from 1858.32 to 898.35 while increasing PSNR from 33.83 to 36.89 and SSIM from 0.8641 to 0.8744.

%\vspace{-5mm}
\begin{table}[b]%[th!b]
    \begin{center}
        \caption{Quantitative Results for Different Geometries.}
        \begin{tabular}{ c c | c c c }
            \hline\hline
            \multicolumn{2}{c}{} & MSE & PSNR & SSIM \\
            \hline
            \multirow{4}{*}{Patient1} & DL-DECT & 2466.07 & 32.37 & 0.8558 \\
            & Parallel & \textbf{866.48} & \textbf{36.84} & \textbf{0.8707} \\
            & Equiangular & 875.06 & 36.80 & 0.8702 \\
            & Equispacing & 871.65 & 36.82 & 0.8703 \\
            \hline
            \multirow{4}{*}{Patient2} & DL-DECT & 1530.75 & 34.61 & 0.8723 \\
            & Parallel & \textbf{876.68} & \textbf{37.03} & \textbf{0.8793} \\
            & Equiangular & 887.60 & 36.99 & 0.8789 \\
            & Equispacing & 882.76 & 37.01 & 0.8789 \\
            \hline
            \multirow{4}{*}{Patient3} & DL-DECT & 1578.13 & 34.61 & 0.8660 \\
            & Parallel & 927.47 & 36.89 & 0.8745 \\
            & Equiangular & 932.40 & 36.88 & 0.8741 \\
            & Equispacing & \textbf{927.15} & \textbf{36.90} & \textbf{0.8747} \\
            \hline\hline
        \end{tabular}
        \label{tab:4}
    \end{center}
\end{table}

\begin{table}[t]%[th!b]
    \begin{center}
        \caption{Computation Time for Different Methods in Seconds.}
        \begin{tabular}{ c c c c }
            \hline\hline
            & Patient1 & Patient2 & Patient3 \\
            \hline
            Slices & 308 & 265 & 387 \\
            \hline
            DL-DECT & 5.43 & 4.68 & 6.86 \\
            Parallel & \textbf{2.33} & \textbf{2.04} & \textbf{2.94} \\
            Equiangular & 2.65 & 2.31 & 3.34 \\
            Equispacing & 2.53 & 2.18 & 3.23 \\
            \hline\hline
        \end{tabular}
        \label{tab:5}
        \vspace{-5mm}
    \end{center}
\end{table}

The computation time for different methods are displayed and compared in Table \ref{tab:5}. Note that computation time for the denoising CNN is not included in the results. The denoising network takes about 0.01 seconds for each slice and the denoising time is similar to all methods. Compared to the DL-DECT method, the proposed method speeds-up the computation time by 2-fold which can be attributed to the reduced number of weights and the simple network structure. There are some differences among the time using different geometries which means the computational cost of the proposed method depends on the projector. %Geometries are not selective in clinical use as they are defined by the machine. However, those complex but more accurate projectors will definitely cost more computational time and may achieve better results.

\section{Discussion}
There are substantial redundant information and correlation in both anatomical structure and energy-domain between the low- and high-energy DECT images. By incorporating the redundancy and the correlation into a deep learning model, it’s possible to provide material- and energy-specific images using standard SECT scanners, which has the potential to alleviate the need for premium DECT scanners. In addition, compared to the standard fully low- and high-energy sampling DECT mechanism, the use of sparse sampling at the second energy level can significantly reduce the radiation dose of DECT imaging.

%To date, few studies have been proposed to exploit the spatial consistency of the DECT images, including our previous DL-DECT method\cite{ref46}.
Our results show both the DL-DECT and FLESH-DECT methods achieve high-performance DECT imaging by using the input low-energy CT data, and quantitative analysis shows FLESH-DECT outperforms DL-DECT in terms of HU accuracy and calculation speed. The superior performance of the FLESH-DECT can be attributed to the additional single-view high-energy projection. Different from the DL-DECT method which directly infers a high-energy image using the incorporated prior knowledge, the proposed FLESH-DECT method uses the learned knowledge to fit the measured high-energy projection. Namely, the high-energy projection introduces a penalty to constrain the projection generated by the predicted high-energy images to be consistent with the measurement, which in turn enhances the accuracy of the predicted images. The single-view high-energy projection can be obtained shortly before or after the standard SECT low-energy data acquisition. For example, one may use a prescanning scout-view image (with the corresponding high-energy kV setting) as the single-view high-energy projection, and existing SECT systems are able to implement these scanning protocols without modifying the hardware.

FLESH-DECT is suitable for different geometries. In this study, we have tested the method using 2D geometries (fan-beam and parallel beam). However, the method can be applied straightforwardly to 3D geometries by extending the networks and the projection operators into 3D scenarios. Since the reconstructed size is smaller than the projection view size in the rotation axis for 3D case, projections generated using the reconstructed image volume cannot match the measured projection. To solve this issue, one can use the middle part of the measured projection which does not propagate through the region beyond the image volume.

\begin{figure*}[tb]
    \centering
        \includegraphics[width=0.98\textwidth]{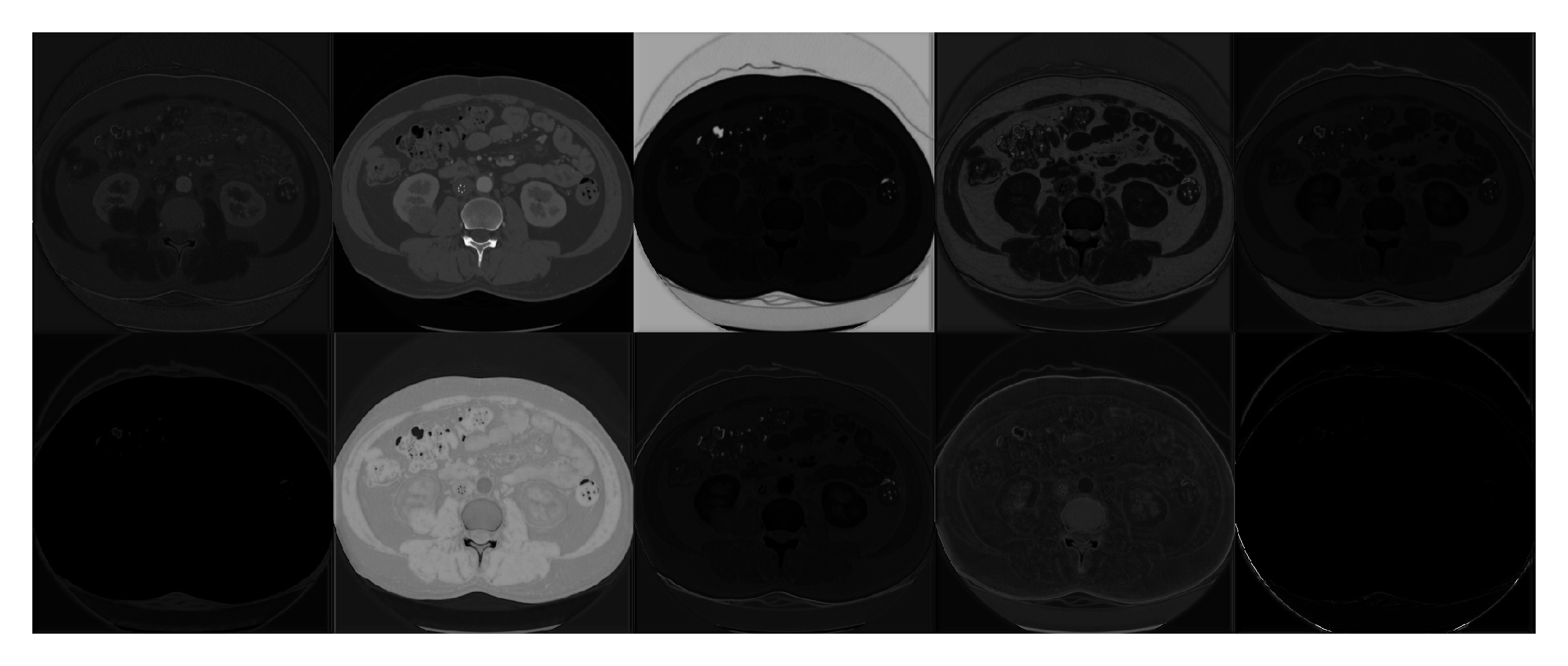}
        \caption{An example of the output of the MD-CNN. Each image represents a channel in the resulting matrix \(A_{DL}\). All images are displayed under the same window with center=0.5 and width=1.0.}
    \label{fig:10}
\end{figure*}

The proposed method employs the MD-CNN to generate “material-decomposition” maps $A$ from 100 kV images. Since basis materials in the image do not change under X-ray at different energy levels, accurate material-decomposition maps are able to provide CT images under any spectrum if combined with the projection view acquired using the corresponding spectrum (e.g. 120 kV image from 120 kV projection, 150 kV image from 150 kV projection). In our experiments, the models were trained and tested using DECT images under 100 kV/Sn 140 kV scanning protocol. Therefore, the generated “material-decomposition” maps are likely to be optimized for this specific protocol and may not be applied to images acquired using a different protocol (such as 80 kV/Sn 140 kV). An example is shown in Fig. \ref{fig:10}. However, if the model was trained using images acquired from several different spectra, the “material-decomposition” maps would be much closer to the real ones and the model would have the potential to generate images under different spectra without re-training the MD-CNN. Also, regularizers may further be introduced and applied to matrix $A$ during network training to enhance the robustness of the model.

There is a denoising-CNN included in the flowchart of FLESH-DECT. We use this network to reduce the impact of image noise and the results show its effectiveness. However, the denoising network is not mandatory and it can be replaced with other image denoising techniques~\cite{ma2011low}, such as non-local mean (NLM)~\cite{zhang2013iterative}, block-matching and 3D filtering (BM3D)~\cite{salehjahromi2017spectral}, without seeing severe degradation in performance. The denoising step may also be removed when the inputting extremely-high quality images.

Despite all the advantages and potentials mentioned above, the proposed method has limitations. FLESH-DECT relies highly on the deep neural network to perform the material-decomposition-like operation. Since the domain knowledge learned by the network highly depends on training data, it is unlikely to provide reasonable results when there is a huge difference between training and testing data. For example, %models trained on abdomen images only may not perform well on brain images or cardiac images,
models trained under 100 kV/Sn 140 kV protocol may not generate correct results when inputting low-energy images scanned under 70 kV protocol. However, these limitations can be solved by training different models for different protocols.

\section{Conclusion}
In this paper, we proposed a deep learning approach to perform DECT imaging using a low-energy image and a single-view high-energy projection. Compared to the standard DECT imaging, the approach can provide superior material-specific images with significantly reduced noise. It also has the potential to simplify the system design and reduce the radiation dose, and allows us to perform high-quality DECT imaging without the conventional hardware-based DECT solutions. The approach may significantly extend the usage of the widespread standard SECT scanners by providing advanced DECT clinical applications, such as urinary stone characterization and as differentiating intracerebral hemorrhage from iodinated contrast, and thus lead to a new paradigm of SECT imaging. % reconstruction quality.

\bibliographystyle{IEEEtran}

%\bibliography{IEEEabrv,paper_v0}

\begin{thebibliography}{10}

\bibitem{ref1}
C.~H. McCollough, S.~Leng, L.~Yu, and J.~G. Fletcher, ``Dual-and multi-energy
  ct: principles, technical approaches, and clinical applications,''
  \emph{Radiology}, vol. 276, no.~3, pp. 637--653, 2015.

\bibitem{ref2}
R.~E. Alvarez and A.~Macovski, ``Energy-selective reconstructions in x-ray
  computerised tomography,'' \emph{Physics in Medicine \& Biology}, vol.~21,
  no.~5, p. 733, 1976.

\bibitem{ref3}
W.~A. Kalender, W.~Perman, J.~Vetter, and E.~Klotz, ``Evaluation of a prototype
  dual-energy computed tomographic apparatus. i. phantom studies,''
  \emph{Medical physics}, vol.~13, no.~3, pp. 334--339, 1986.

\bibitem{ref4}
T.~G. Flohr \emph{et~al.}, ``First performance evaluation of a dual-source ct
  (dsct) system,'' \emph{European radiology}, vol.~16, no.~2, pp. 256--268,
  2006.

\bibitem{ref5}
T.~R. Johnson \emph{et~al.}, ``Material differentiation by dual energy ct:
  initial experience,'' \emph{European radiology}, vol.~17, no.~6, pp.
  1510--1517, 2007.

\bibitem{ref6}
D.~T. Boll, E.~M. Merkle, E.~K. Paulson, R.~A. Mirza, and T.~R. Fleiter,
  ``Calcified vascular plaque specimens: assessment with cardiac dual-energy
  multidetector ct in anthropomorphically moving heart phantom,''
  \emph{Radiology}, vol. 249, no.~1, pp. 119--126, 2008.

\bibitem{lee2017feasibility}
D.~Lee \emph{et~al.}, ``A feasibility study of low-dose single-scan dual-energy
  cone-beam ct in many-view under-sampling framework,'' \emph{IEEE transactions
  on medical imaging}, vol.~36, no.~12, pp. 2578--2587, 2017.

\bibitem{petrongolo2018single}
M.~Petrongolo and L.~Zhu, ``Single-scan dual-energy ct using primary
  modulation,'' \emph{IEEE transactions on medical imaging}, vol.~37, no.~8,
  pp. 1799--1808, 2018.

\bibitem{xue2019accurate}
Y.~Xue \emph{et~al.}, ``Accurate multi-material decomposition in dual-energy
  ct: A phantom study,'' \emph{IEEE Transactions on Computational Imaging},
  vol.~5, no.~4, pp. 515--529, 2019.

\bibitem{ref7}
L.~Yu, S.~Leng, and C.~H. McCollough, ``Dual-energy ct--based monochromatic
  imaging,'' \emph{American journal of Roentgenology}, vol. 199, no.
  5\_supplement, pp. S9--S15, 2012.

\bibitem{ref8}
S.~R. Pomerantz \emph{et~al.}, ``Virtual monochromatic reconstruction of
  dual-energy unenhanced head ct at 65--75 kev maximizes image quality compared
  with conventional polychromatic ct,'' \emph{Radiology}, vol. 266, no.~1, pp.
  318--325, 2013.

\bibitem{ref9}
C.~Phan, A.~Yoo, J.~Hirsch, R.~Nogueira, and R.~Gupta, ``Differentiation of
  hemorrhage from iodinated contrast in different intracranial compartments
  using dual-energy head ct,'' \emph{American journal of neuroradiology},
  vol.~33, no.~6, pp. 1088--1094, 2012.

\bibitem{ref10}
W.~H. Sommer \emph{et~al.}, ``The value of dual-energy bone removal in maximum
  intensity projections of lower extremity computed tomography angiography,''
  \emph{Investigative radiology}, vol.~44, no.~5, pp. 285--292, 2009.

\bibitem{ref11}
B.~Buerke, G.~Wittkamp, H.~Seifarth, W.~Heindel, and S.~P. Kloska,
  ``Dual-energy cta with bone removal for transcranial arteries:
  intraindividual comparison with standard cta without bone removal and
  tof-mra,'' \emph{Academic radiology}, vol.~16, no.~11, pp. 1348--1355, 2009.

\bibitem{ref12}
D.~Morhard, C.~Fink, A.~Graser, M.~F. Reiser, C.~Becker, and T.~R. Johnson,
  ``Cervical and cranial computed tomographic angiography with automated bone
  removal: dual energy computed tomography versus standard computed
  tomography,'' \emph{Investigative radiology}, vol.~44, no.~5, pp. 293--297,
  2009.

\bibitem{ref13}
B.~Schulz \emph{et~al.}, ``Automatic bone removal technique in whole-body
  dual-energy ct angiography: performance and image quality,'' \emph{American
  Journal of Roentgenology}, vol. 199, no.~5, pp. W646--W650, 2012.

\bibitem{ref14}
N.~Takahashi \emph{et~al.}, ``Dual-energy ct iodine-subtraction virtual
  unenhanced technique to detect urinary stones in an iodine-filled collecting
  system: a phantom study,'' \emph{American Journal of Roentgenology}, vol.
  190, no.~5, pp. 1169--1173, 2008.

\bibitem{ref15}
J.~Ferda \emph{et~al.}, ``The assessment of intracranial bleeding with virtual
  unenhanced imaging by means of dual-energy ct angiography,'' \emph{European
  radiology}, vol.~19, no.~10, pp. 2518--2522, 2009.

\bibitem{ref16}
A.~Graser \emph{et~al.}, ``Dual-energy ct in patients suspected of having renal
  masses: can virtual nonenhanced images replace true nonenhanced images?''
  \emph{Radiology}, vol. 252, no.~2, pp. 433--440, 2009.

\bibitem{ref17}
L.~M. Ho \emph{et~al.}, ``Characterization of adrenal nodules with dual-energy
  ct: can virtual unenhanced attenuation values replace true unenhanced
  attenuation values?'' \emph{American Journal of Roentgenology}, vol. 198,
  no.~4, pp. 840--845, 2012.

\bibitem{ref18}
S.~Mangold \emph{et~al.}, ``Virtual nonenhanced dual-energy ct urography with
  tin-filter technology: determinants of detection of urinary calculi in the
  renal collecting system,'' \emph{Radiology}, vol. 264, no.~1, pp. 119--125,
  2012.

\bibitem{ref19}
M.~Toepker \emph{et~al.}, ``Virtual non-contrast in second-generation,
  dual-energy computed tomography: reliability of attenuation values,''
  \emph{European journal of radiology}, vol.~81, no.~3, pp. e398--e405, 2012.

\bibitem{ref20}
A.~N. Primak \emph{et~al.}, ``Noninvasive differentiation of uric acid versus
  non--uric acid kidney stones using dual-energy ct,'' \emph{Academic
  radiology}, vol.~14, no.~12, pp. 1441--1447, 2007.

\bibitem{ref21}
D.~T. Boll \emph{et~al.}, ``Renal stone assessment with dual-energy
  multidetector ct and advanced postprocessing techniques: improved
  characterization of renal stone composition鈥攑ilot study,''
  \emph{Radiology}, vol. 250, no.~3, pp. 813--820, 2009.

\bibitem{ref22}
G.~Ascenti \emph{et~al.}, ``Stone-targeted dual-energy ct: a new diagnostic
  approach to urinary calculosis,'' \emph{American Journal of Roentgenology},
  vol. 195, no.~4, pp. 953--958, 2010.

\bibitem{ref25}
S.~Leng \emph{et~al.}, ``Feasibility of discriminating uric acid from non--uric
  acid renal stones using consecutive spatially registered low-and high-energy
  scans obtained on a conventional ct scanner,'' \emph{American Journal of
  Roentgenology}, vol. 204, no.~1, pp. 92--97, 2015.

\bibitem{ref26}
S.~Leng, L.~Yu, J.~G. Fletcher, and C.~H. McCollough, ``Maximizing iodine
  contrast-to-noise ratios in abdominal ct imaging through use of energy domain
  noise reduction and virtual monoenergetic dual-energy ct,'' \emph{Radiology},
  vol. 276, no.~2, pp. 562--570, 2015.

\bibitem{ref27}
W.~Zhao \emph{et~al.}, ``Using edge-preserving algorithm with non-local mean
  for significantly improved image-domain material decomposition in dual-energy
  ct,'' \emph{Physics in Medicine \& Biology}, vol.~61, no.~3, p. 1332, 2016.

\bibitem{ref28}
V.~Gulshan \emph{et~al.}, ``Development and validation of a deep learning
  algorithm for detection of diabetic retinopathy in retinal fundus
  photographs,'' \emph{Jama}, vol. 316, no.~22, pp. 2402--2410, 2016.

\bibitem{ref29}
A.~Esteva \emph{et~al.}, ``Dermatologist-level classification of skin cancer
  with deep neural networks,'' \emph{Nature}, vol. 542, no. 7639, pp. 115--118,
  2017.

\bibitem{ref30}
D.~S.~W. Ting \emph{et~al.}, ``Development and validation of a deep learning
  system for diabetic retinopathy and related eye diseases using retinal images
  from multiethnic populations with diabetes,'' \emph{Jama}, vol. 318, no.~22,
  pp. 2211--2223, 2017.

\bibitem{ref31}
F.~Liu, H.~Jang, R.~Kijowski, T.~Bradshaw, and A.~B. McMillan, ``Deep learning
  mr imaging--based attenuation correction for pet/mr imaging,''
  \emph{Radiology}, vol. 286, no.~2, pp. 676--684, 2018.

\bibitem{ref32}
L.~Xing, E.~A. Krupinski, and J.~Cai, ``Artificial intelligence will soon
  change the landscape of medical physics research and practice,''
  \emph{Medical physics}, vol.~45, no.~5, pp. 1791--1793, 2018.

\bibitem{ref33}
W.~Zhao \emph{et~al.}, ``Markerless pancreatic tumor target localization
  enabled by deep learning,'' \emph{International Journal of Radiation
  Oncology* Biology* Physics}, vol. 105, no.~2, pp. 432--439, 2019.

\bibitem{ref34}
A.~K. Maier \emph{et~al.}, ``Learning with known operators reduces maximum
  error bounds,'' \emph{Nature machine intelligence}, vol.~1, no.~8, pp.
  373--380, 2019.

\bibitem{ref35}
L.~Shen, W.~Zhao, and L.~Xing, ``Patient-specific reconstruction of volumetric
  computed tomography images from a single projection view via deep learning,''
  \emph{Nature biomedical engineering}, vol.~3, no.~11, pp. 880--888, 2019.

\bibitem{ref36}
D.~Lee, H.~Kim, B.~Choi, and H.-J. Kim, ``Development of a deep neural network
  for generating synthetic dual-energy chest x-ray images with single x-ray
  exposure,'' \emph{Physics in Medicine \& Biology}, vol.~64, no.~11, p.
  115017, 2019.

\bibitem{ref37}
W.~Zhao \emph{et~al.}, ``Incorporating imaging information from deep neural
  network layers into image guided radiation therapy (igrt),''
  \emph{Radiotherapy and Oncology}, vol. 140, pp. 167--174, 2019.

\bibitem{ref38}
H.~Chen \emph{et~al.}, ``Low-dose ct with a residual encoder-decoder
  convolutional neural network,'' \emph{IEEE transactions on medical imaging},
  vol.~36, no.~12, pp. 2524--2535, 2017.

\bibitem{ref39}
Q.~Yang \emph{et~al.}, ``Low-dose ct image denoising using a generative
  adversarial network with wasserstein distance and perceptual loss,''
  \emph{IEEE transactions on medical imaging}, vol.~37, no.~6, pp. 1348--1357,
  2018.

\bibitem{ref40}
J.~M. Wolterink, T.~Leiner, M.~A. Viergever, and I.~I{\v{s}}gum, ``Generative
  adversarial networks for noise reduction in low-dose ct,'' \emph{IEEE
  transactions on medical imaging}, vol.~36, no.~12, pp. 2536--2545, 2017.

\bibitem{ref41}
E.~Kang, W.~Chang, J.~Yoo, and J.~C. Ye, ``Deep convolutional framelet denosing
  for low-dose ct via wavelet residual network,'' \emph{IEEE transactions on
  medical imaging}, vol.~37, no.~6, pp. 1358--1369, 2018.

\bibitem{ref42}
Y.~Liao \emph{et~al.}, ``Pseudo dual energy ct imaging using deep
  learning-based framework: basic material estimation,'' in \emph{Medical
  Imaging 2018: Physics of Medical Imaging}, vol. 10573.\hskip 1em plus 0.5em
  minus 0.4em\relax International Society for Optics and Photonics, 2018, p.
  105734N.

\bibitem{ref43}
W.~Zhang \emph{et~al.}, ``Image domain dual material decomposition for
  dual-energy ct using butterfly network,'' \emph{Medical physics}, vol.~46,
  no.~5, pp. 2037--2051, 2019.

\bibitem{ref44}
M.~G. Poirot \emph{et~al.}, ``physics-informed deep learning for dual-energy
  computed tomography image processing,'' \emph{Scientific reports}, vol.~9,
  no.~1, pp. 1--9, 2019.

\bibitem{ref45}
C.~Feng, K.~Kang, and Y.~Xing, ``Fully connected neural network for virtual
  monochromatic imaging in spectral computed tomography,'' \emph{Journal of
  Medical Imaging}, vol.~6, no.~1, p. 011006, 2018.

\bibitem{ref46}
W.~Zhao, T.~Lv, R.~Lee, Y.~Chen, and L.~Xing, ``Obtaining dual-energy computed
  tomography (ct) information from a single-energy ct image for quantitative
  imaging analysis of living subjects by using deep learning,'' in \emph{Pac
  Symp Biocomput}.\hskip 1em plus 0.5em minus 0.4em\relax World Scientific,
  2020.

\bibitem{ref50}
R.~E. Alvarez and A.~Macovski, ``Energy-selective reconstructions in x-ray
  computerised tomography,'' \emph{Physics in Medicine \& Biology}, vol.~21,
  no.~5, p. 733, 1976.

\bibitem{ref51}
O.~Ronneberger, P.~Fischer, and T.~Brox, ``U-net: Convolutional networks for
  biomedical image segmentation,'' in \emph{International Conference on Medical
  image computing and computer-assisted intervention}.\hskip 1em plus 0.5em
  minus 0.4em\relax Springer, 2015, pp. 234--241.

\bibitem{ref47}
A.~Vedaldi and K.~Lenc, ``Matconvnet: Convolutional neural networks for
  matlab,'' in \emph{Proceedings of the 23rd ACM international conference on
  Multimedia}, 2015, pp. 689--692.

\bibitem{ref48}
M.~Abadi \emph{et~al.}, ``Tensorflow: Large-scale machine learning on
  heterogeneous distributed systems,'' \emph{arXiv preprint arXiv:1603.04467},
  2016.

\bibitem{ref49}
D.~P. Kingma and J.~Ba, ``Adam: A method for stochastic optimization,''
  \emph{arXiv preprint arXiv:1412.6980}, 2014.

\bibitem{ma2011low}
J.~Ma \emph{et~al.}, ``Low-dose computed tomography image restoration using
  previous normal-dose scan,'' \emph{Medical physics}, vol.~38, no.~10, pp.
  5713--5731, 2011.

\bibitem{zhang2013iterative}
H.~Zhang \emph{et~al.}, ``Iterative reconstruction for x-ray computed
  tomography using prior-image induced nonlocal regularization,'' \emph{IEEE
  Transactions on Biomedical Engineering}, vol.~61, no.~9, pp. 2367--2378,
  2013.

\bibitem{salehjahromi2017spectral}
M.~Salehjahromi, Y.~Zhang, and H.~Yu, ``A spectral ct denoising algorithm based
  on weighted block matching 3d filtering,'' in \emph{Developments in X-Ray
  Tomography XI}, vol. 10391.\hskip 1em plus 0.5em minus 0.4em\relax
  International Society for Optics and Photonics, 2017, p. 103910G.

\end{thebibliography}

\end{document}